\documentclass[12pt]{article}
\usepackage{epsfig}

\usepackage{amsthm}
\usepackage{amsfonts}
\usepackage{amssymb}
\usepackage{graphicx}
\usepackage{amsmath}
\usepackage{lscape}
\usepackage{lscape,color}

\definecolor{c30}{rgb}{0,.7,.2}

\definecolor{d30}{rgb}{1,0,0}

\newcommand{\cO}{\mathcal{O}}

\newcommand{\bea}{\begin{eqnarray}}
\newcommand{\eea}{\end{eqnarray}}
\newcommand{\nn}{\nonumber}
\newcommand{\be}{\begin{equation}}
\newcommand{\ee}{\end{equation}}
\newcommand{\qu}{\quad}
\renewcommand{\t}{\tau}
\newcommand{\tb}{\bar{\tau}}
\newcommand{\pa}{\partial}
\newcommand{\tpt}{\tau\frac{\partial}{\partial\tau}}
\newcommand{\btpt}{\bar{\tau}\frac{\partial}{\partial\bar{\tau}}}

\newtheorem{Theorem}{Theorem}[section]

\newtheorem{Proof}[Theorem]{Proof}

\newtheorem{Lemma}[Theorem]{Lemma}

\begin{document}
\begin{small}
\title{Chiral Modular Bootstrap  }
\author{
M. Ashrafi\thanks{E-mail address: mashrafi@ucdavis.edu}\\[6pt]
 Department of Physics,University of California Davis \\One Shields Avenue, Davis, California 95616, USA }
\date{}
\maketitle
\begin{abstract}
Using modular bootstrap we show the lightest primary fields of a unitary compact two dimensional conformal field theory(with $c, \bar{c}>1$)  has a conformal  weight  $h_1\le \frac{c}{12}+\cO(1)$.This  implies that the upper bound on the dimension of the lightest primary fields depends on their spin. In particular if the set of lightest primary fields includes extremal or near extremal states whose spin to dimension ratio $\frac{j}{\Delta}\approx 1$, the corresponding dimension is $\Delta\le \frac{c}{12}+\cO(1)$. From AdS/CFT correspondence, we obtain  an upper bound on the spectrum of black hole in  three dimensional gravity. Our results show that if the first primary fields have large spin, the corresponding three dimensional gravity has extremal or near extremal BTZ black hole.     \\

\end{abstract}
\end {small}

\section{Introduction}
Modular bootstrap is a powerful approach to study the spectrum of unitary compact two dimensional CFTs (UC-CFT$_2$) which is encapsulated in the  conformal weights $(h,\bar h)$ of the primary fields and their degeneracy $\rho(h,\bar h)$. $h$ and $\bar h$ assume discrete values and  are non-negative. The  vacuum state $(0,0)$ is unique, i.e. $\rho(0,0)=1$, and it is invariant under  global conformal transformations. Any two dimensional CFT enjoys a conformal symmetry whose generators sit in two copies of the virasoro algebra.  These data are encoded in the following partition function
    \be
     Z(\tau,\bar \tau)=\sum_{h,\bar h}\chi_h(\tau)\bar \chi_{\bar h}(\bar\tau),
    \ee
where $\chi(\t)$ is virasoro character\footnote{For recent development in modular bootstrap see \cite{Maxfield:2019hdt}-\cite{Dyer:2017rul}}.

It is supposed that UC-CFT$_2$s  have no extended chiral algebras beyond the virasoro algebra and consequently no primary fields with $h\bar h=0$. In such models, modular bootstrap initiated in \cite{Hellerman:2009bu} puts an upper bound on the dimension $\Delta:=h+\bar h$ of the lightest primary fields $\Delta<\frac{c}{6}+\cO(1/c)$. This upper bound has been obtained by studying the invariance of the partition function $Z(\tau,\bar \tau)$ under the $S$-transformation $(\tau,\bar\tau)\to(-\tau^{-1},-{\bar\tau}^{-1})$ around the self-dual point $(\tau,\bar \tau)=(i,-i)$ which implies that
    \begin{align}\label{mbs-general}
    &\left.(\tau\partial_\tau)^{N_L}(\bar\tau\partial_{\bar{\tau}})^{N_R}Z(\tau,\bar\tau)\right|_{\tau=-\bar\tau=i}=0
    &\mbox{for}\ \ \ N_L+N_R=\mbox{odd}.
    \end{align}
 In  \cite{Hellerman:2009bu, Keller:2013qqa, Friedan:2013cba, Qualls:2013eha},  the authors investigated this constraint for $(\tau,\bar\tau)=(\frac{i\beta}{2\pi} ,-\frac{i\beta}{2\pi})$ with $\beta\in\mathbb{R}$ and obtained the upper-bound mentioned above. In this approach the partition function $Z(\beta):=Z(\frac{i\beta}{2\pi},-\frac{i\beta}{2\pi})$ is independent of the spin of the primary fields given by $j:=h-\bar h$ and depends only on their dimensions $\Delta$, and Eq.\eqref{mbs-general} reads
    \begin{align}\label{mbs-beta}
    &(\beta\partial_\beta)^NZ(\beta)=0&\mbox{for}\ \ \ N=\mbox{odd}.
    \end{align}
    
 There are also  UC-CFT$_2$s with primary fields such that $h\bar h=0$. The principal examples are the extremal CFT's. For extremal CFTs the scaling dimension of the first primary field after vacuum state is $\frac{c}{24}+1$ \cite{hon1, hon2} . For extremal CFTs which are holomorphicaly factorizable, the upper bound on the first primary field is $\Delta< {\rm min}(\frac{c}{24}+1,\frac{\bar{c}}{24}+1)$\cite{witten}. From AdS/CFT correspondence, extremal CFT are suppose to be dual to pure gravity\cite{witten, Maloney:2007ud}. The first primary fields above the vacuum  correspond with a BTZ black hole. 
 
 All known theories of gravity contain BTZ black hole. Since the lightest primary fields corresponding to BTZ black hole are   $\frac{c}{12}$, the upper bound  obtained from  modular bootstrap is weaker than what we expected from holography. 
 
Additionally, there is another class of CFTs , half integer CFT (HI-CFT), where  upper bound is $\Delta=\frac{c}{12}$\cite{Ashrafi-16}. HI-CFTs are  CFTs which are not holomorphicaly factorizable and the scaling dimension of primary fields are half integer. It is also known  that in ${\mathcal N} = (1, 1)$ supersymmetric theories with a $U(1)$ current we have $\Delta<\frac{c}{12}+\cO(1/c)$ \cite{Kachru-16}, which is also the case in UC-CFT$_2$'s having only primaries  whose spin $s$ are even \cite{Qualls-14}.
Therefore we expect that the upper bound should be improved. 

Recently,  numerical methods have lowered the upper bound to  $\frac{c}{9}$\cite{Lin-16, Afkhami-Jeddi:2019zci}.  An important observation is that in Eq.\eqref{mbs-beta}, the limits $c\to\infty$ and $N\to \infty$, do not commute with each other \cite{Lin-16}. In this paper we prove that the upper bound obtained in \cite{Hellerman:2009bu} by considering $N=1,3$ cannot be improved by trying any $N<\sqrt c$. This would require going to $N>c$ as has been done numerically in \cite{Lin-16, Afkhami-Jeddi:2019zci}. However, there is  limitations on the central charges in the numerical method. The modular bootstrap extended up to $c\sim 1800$.
 
 Most of the recent work on modular bootstrap is limited to self dual point $\t=i$, which partition function  $(\tau,\bar\tau)=(\frac{i\beta}{2\pi},-\frac{i\beta}{2\pi})$ is independent of spin, and the theory does not have any chiral algebra beyond the virasoro algebra.   In this paper we released these two constraints and studied the modular bootstrap in grand canonical ensemble. We obtain an upper bound on the conformal weight of the lightest primary fields
    \begin{align}\label{upper-bound}
    &h_1\le \frac{c}{12}+\cO(1),& \bar h_1\le  \frac{\bar{c}}{12}+\cO(1).
    \end{align}
 Our results are compatible with the upper bound on the conformal weight of the lightest primary fields of  CFT's invariant under the parity transformation obtained in \cite{Shaghoulian-18}.   We will show that the upper bound \eqref{upper-bound} can not be improved by studying $N_L,N_R<\sqrt c$. This leaves room for  investigating Eq.\eqref{mbs-general} for, say, $N_L>c$ in search of an upper bound smaller than \eqref{upper-bound}.
 
 This paper is organized as follows. First, we  review the HI-CFT. Holomorphicaly factorizable partition function are a subclass of this class of CFT.  Next, we obtain the  upper bound on the first primary field. In section \ref{partition function}, we  derive the closed form for the $N$ order derivative of the partition function. In section \ref{secch}, we  consider the  grand canonical partition function for CFTs which can also have chiral algebra beyond virasoro algebra. Using the medium temperature expansion for the chiral and anti-chiral sector of the theory, separately we  derive a spin dependent upper bound on the primary fields. We show that for CFTs which lowest primary fields has heavy spin, the upper bound  improves as we expect from holography. In the section \ref{even spin} we use the medium temperature expansion with intermediate temperature expansion to determine the spin dependent upper bound for even spin CFTs . Finally, section \ref{summery} summarizes our results.
 \section{Review HI-CFT upper Bound}\label{review}
 In \cite{Ashrafi-16}, it is shown that corresponding to every $S$ invariant partition function $Z(\t,\tb)$, there is a chiral $S$ invariant partition function ${\cal Z}(\tau):=Z(\tau,-\tau)$
  \begin{align}\label{chim}
    \mathit{ch} : Z(\t , \tb)\rightarrow{\cal Z}({\t} ),
 \end{align}
 which is called $  \mathit{ch}$ image of partition function $Z(\t,\tb)$ where
 \be\label{z3}
    {\cal Z}(\tau)=q^{-\frac{c_{\rm tot}}{24}}\sum_{\Delta=0}\hat\rho(\Delta)q^\Delta,
    \ee
    and
    \be
    \hat \rho(\Delta):=\sum_{j\in {\cal J}_\Delta}\rho(\Delta,j).
    \ee
There is a family of CFTs  in which $\Delta$  is an integer number .The corresponding conformal weights $h,\bar{h}$ are half-integers. Such a CFT  called an HI-CFT. It was shown that the total central charge $c_{tot}=c+\bar{c}$ for HI-CFT is an integer multiple of $8$, $c_{tot}=8k, k\in {\mathbb N}$. Therefor, right and left central charges are integer multiple of 4:
       \begin{align}
      \label{c2}
 &      c\in\,4\,{\mathbb N}, & \bar{c}\in 4\,{\mathbb N}.
      \end{align}

In \cite{Ashrafi-16}, it was shown that the $  \mathit{ch}$ image of partition function for HI-CFT  has expansion in terms of $ \mathfrak{j} (\t)$ as follows
 \begin{align}
     \label{ch1}
    &{\cal Z}(\t)=\mathfrak{j}^{k}\sum_{r=0}^{[k/3]}n_r J^{-r}, &n_r\in {\mathbb N}.
    \end{align}
     where the function $\mathfrak{j}$ has expansion as follows
 \begin{align}
    \label{g11}
    \mathfrak{j} (\t)&:=\frac{1}{2}\left[ \left(\sqrt{\frac{\theta_2(\t)}{\eta(\t)}}\right)^{16}+\left(\sqrt{\frac{\theta_3(\t)}{\eta(\t)}}\right)^{16}+\left(\sqrt{\frac{\theta_4(\t)}{\eta(\t)}}\right)^{16}\right] \nn\\
     &=q^{\frac{-1}{3}}\left(1+248\,q+\cdots\right).
     \end{align}
         In order to find the extremal  partition function ${\cal Z}(\t)$, we should set the coefficients  $n_r$ in such a way that the terms with $q^{-n}$, for $n=0,....,k$ in ${\cal Z}(\t)$ coincide with the corresponding terms in the $  \mathit{ch}$ image of vacuum partition function ${\cal Z}^0(\t)$. Therefore, the upper bound on the scaling dimension of the first primary fields for HI-CFT can be obtained as follows
      \be\label{uphicft}
    \Delta_1=\left[\frac{c_{\rm tot}}{24}\right]+1.
      \ee
     Some examples of extremal HI-CFTs with $c_{tot}=8,16,32,48$ are
      \be
    { \cal Z}_8(\t ) =\mathfrak{j}(\t ) = q^{-1/3} + 248 q^{2/3} + 4124 q^{5/3} +  . . . ,
      \ee 
       \be
    { \cal Z}_{16}(\t ) =\mathfrak{j}^{2}(\t ) = q^{-2/3} + 496 q^{1/3} + 69752 q^{4/3} . . .,
          \ee
              \be
    { \cal Z}_{32}(\t ) =\mathfrak{j}^{4}(\t )-992\mathfrak{j} = q^{-4/3} + 139504 q^{2/3} + 69332992 q^{5/3} +. . . ,
      \ee
       \be
    { \cal Z}_{48}(\t ) =\mathfrak{j}^{5}(\t )-1240\mathfrak{j}^2 = q^{-5/3} + 20620 q^{1/3} + 86666240 q^{4/3} + 24243884350 q^{7/3} +  . . . .
      \ee
      
\section{Partition Function}\label{partition function}
The partition function of unitary two dimensional CFT on a torus with complex structure $\tau=\tau_1+i\tau_2$, is defined as follows
 \begin{align}\label{p1}
Z(\tau,\bar{\tau})={\rm Tr}\,\left(e^{2\pi i\tau (L_0-\frac{c}{24})}e^{-2\pi i\bar{\tau} (\bar{L}_0-\frac{\bar{c}}{24})}\right),
\end{align}
where $\bar{\tau}=\tau_1-i\tau_2$, is the complex conjugate of $\tau$, $c$ and $\bar{c}$ are left and right central charges, respectively. 

For CFTs with $c,\bar{c}>1$,  the partition function can be written in terms of Virasoro character $\chi _h(\tau)(\bar{\chi} _{\bar{h}}(\bar{\tau}))$ as follows
\begin{align}\label{p2}
Z(\tau,\bar{\tau})=\chi _0(\tau)\bar{\chi} _{0}(\bar{\tau})+\sum _{h}\rho(h)\chi _h(\tau)\bar{\chi} _{0}(\bar{\tau})+\sum _{\bar{h}}\rho(\bar{h})\chi _0(\tau)\bar{\chi} _{\bar{h}}(\bar{\tau})+\sum _{h,\bar{h}}\rho(h,\bar{h})\chi _h(\tau)\bar{\chi} _{\bar{h}}(\bar{\tau}),
\end{align}
 where the summation is over all the primary fields, and
\begin{align}\label{ch1}
\chi_h(\tau)=\frac{1}{\eta(\tau)}q^{h+E_0}(1-q)^{\delta _{h,0}},
\end{align}
\begin{align}\label{ch2}
\bar{\chi}_{\bar{h}}(\bar{\tau})=\frac{1}{\bar{\eta}(\bar{\tau})}q^{\bar{h}+\bar{E}_0}(1-\bar{q})^{\delta _{\bar{h},0}}.
\end{align}
 $E_0=\frac{1-c}{24}$, ($\bar{E}_0=\frac{1-\bar{c}}{24}$). $\eta(\tau)$ is Dedekind eta function. 
The states with conformal weight $h_0=0, \bar{h}_0=0$  correspond to the vacuum states  and the states with conformal weight $h_0=0, \bar{h}_0\neq 0$ and $h\neq 0, \bar{h}=0$ indicate  the states with chiral symmetry. $\rho(h)$ and $\rho(\bar{h})$  are the density of holomorphic and antiholomorphic current of spin $h$ and $\bar{h}$ respectively, and $\rho(h,\bar{h})$ is the density of state of primary operators of weight $(h,\bar{h})$.  %
   \subsection{Invariance of partition function under modular transformations}
    The partition function on a torus is invariant under modular transformations. $T$ invariance demand that the spin $j$ should be an integer number $j\in{\mathbb Z}$, and $c-\bar{c}\in24{\mathbb Z}$ \cite{Ashrafi-16}. 
    
      The $S:=\tau\rightarrow\frac{-1}{\tau}, \bar{\tau}\rightarrow\frac{-1}{\bar{\tau}}$  invariance of partition function leads:    
        \begin{equation}\label{z1}
    Z(\t , \tb)=Z\left(-\frac{1}{\t}, -\frac{1}{\tb}\right),
    \end{equation}
    taking the derivative of  both sides of the equation  leads to
       \be\label{m1}
    \left(\tpt\right)^{N_L}\left(\btpt\right)^{N_R}Z(\t , \tb)=(-1)^{N_L+N_R}\left(\omega \frac{\pa}{\pa\omega }\right)^{N_L} \left(\bar{\omega}\frac{\pa}{\pa\bar{\omega}}\right)^{N_R}Z\left(\omega, \bar\omega\right),
    \ee
    where
    \begin{align}
    &\omega := -\frac{1}{\t},&\bar\omega:=-\frac{1}{\bar\tau}.
    \end{align}
 Using \eqref{m1} at the self dual point, leads to the set of constraints on the partition function as follows
  \be\label{de1}
    {\hat D_L}^{N_L}{\hat D_R}^{N_R}Z(\t , \tb)\bigg\vert _{\tau =+i,\tb =-i}=0 \qu \mbox{for}\qu N_L+N_R=\mbox{odd},
    \ee
    where ${\hat D_L}=\tpt$, and ${\hat D_R}=\btpt$ are the left dilatation operator  and the right dilatation operator,  respectively.
    The set of constraints \eqref{de1}, are called medium temperature expansion\cite{Hellerman:2009bu}.

   Now, we consider the invariance of the partition function under   $ST$ transformation. Under this transformation the parameter $\tau$ change as follows \cite{Qualls-14}
\begin{align}\label{es1}
ST:\tau\rightarrow\frac{-1}{\tau +1}.
\end{align}
where self dual point of this transformation is $\frac{-1}{2}+i\frac{\sqrt{3}}{2}$.\\

Using invariance of partition function under $ST$ transformation 
\begin{align}\label{es2}
Z(\tau , \bar{\tau})=Z(-\frac{1}{\tau +1}, -\frac{1}{\bar{\tau} +1}),
\end{align}
and taking the derivative of both side of \eqref{es2} with respect to $\tau$  and $\bar{\tau}$ leads to these constraints :
 \be\label{de11}
    {\hat D_L}^{N_L}{\hat D_R}^{N_R}Z(\t , \tb)\bigg\vert _{\tau =-\tb =\frac{-1}{2}+i\frac{\sqrt{3}}{2}}=0 \qu \mbox{for}\qu N_L \mbox{mod 3}\ne N_R \mbox{mod 3},
    \ee
    which is called the intermediate temperature expansion \cite{Qualls-14}.  
   
 %
 \subsubsection{ $N$th order derivative of  Virasoro character}
In this section we will derive the $N$-th order derivative  of the virasoro character.
\begin{Lemma}\label{lem1}
$N$th order derivative of the virasoro character can be obtained as follow
\begin{align}\label{d1}
\hat{D}_L ^{N_{L}}\chi_h(\tau)= \sum_{n=0}^{N} A_n^{(N)}(\tau)( \tau B_h(\tau))^n\chi_{h}(\tau),
\end{align}
where
\begin{align}
\label{d22-2}
 B_h(\tau)=2\pi i( h+E_0)-\frac{\eta '(\tau)}{\eta (\tau)}-\frac{2\pi i\delta_{h,0}}{e^{-2\pi i\tau}-1}.
\end{align}
and
\begin{itemize}
    \item  $A_n^{(N)}(\tau)=0$ for $n>N $ ,
    \item $A_N^{(N)}(\tau)=1$ ,
    \item  for $n<N $ 
    \begin{small}
\begin{align}\label{A11}
A_{n}^{(N)}(\tau)=\tau\frac{\partial A_{n}^{(N-1)}(\t)}{\partial\tau}+(n+1)\left(\tau^2\frac{\partial}{\partial\tau}B_h(\t)\right)A_{n+1}^{(N-1)}(\t)+nA_{n}^{(N-1)}(\t)+A_{n-1}^{(N-1)}(\t).
\end{align}
\end{small}
    \end{itemize}
\end{Lemma}
\begin{Proof}
We prove the lemma by induction. For $N_L=1$ we have:
\begin{align}
\label{d22-1}
\hat{D}_L \chi_{h}(\tau)=\tau B_h(\tau)\chi_{h}(\tau),
\end{align}
which is obtained with the taking the derivative of the virasoro character. Now we assume \eqref{d1} is correct for $N_L=k-1$, then we show that Eq \eqref{d1} is true  for $N_L=k$.
For $N_L=k-1$:
\begin{align}\label{d22-4}
\hat{D}_L ^{N-1}\chi_h(\tau)= \sum_{n=0}^{N-1} A_n^{(N-1)}(\tau)( \tau B_h(\tau))^n\chi_{h}(\tau).
\end{align}
By acting the dilatation operator on \eqref{d22-4} and using \eqref{A11}, we see that \eqref{d1} is true.
\end{Proof}
Solving the recurrence relation \eqref{A11} leads to
\be\label{A2}
A_n^{N}=\sum_{k=0}^{n}\left(k+\tpt\right)A_{k}^{N-n-1+k}(\t)+(k+1)\left(\tau^2\frac{\partial}{\partial\tau}B_h(\t)\right)A_{k+1}^{N-n-1+k}(\t),
\ee
Using \eqref{A2}, the first few $A_n^{(N)}:= A_n^{(N)}(i)$s  can be obtained as follows
\begin{align}\label{A3}
A_{N-1}^{(N)}&=\frac{N(N-1)}{2},\nonumber\\
A_{N-2}^{(N)}&=\frac{{(N-1)}^2{(N-2)}^2}{2\times 4}+\frac{(N-1)(N-2)(2N-3)}{2\times 6}+\frac{N(N-1)}{2}\left(\tau^2\frac{\partial}{\partial\tau}B_h(\t)\right).
\end{align}
Using\eqref{d1} the $N$ order derivative of the virasoro character  at self dual point obtain as follows
\begin{align}\label{d3}
\hat{D}_L ^{N}\chi_{h}(i)\bigg\vert_{\tau=i}= {(-1)}^N  g^{(N)}(h+E_0)\chi_{h}(i),
\end{align}
\begin{align}\label{d31}
\hat{D}_L ^{N}\chi_{0}(i)\bigg\vert_{\tau=i}={(-1)}^N g^{(N)}(E_0)\chi_{0}(i),
\end{align}
where the polynomial $g^{(N)}(h)$ is defined as below
\begin{align}\label{g1-1}
g^{(N)}(h+E_{0}):=\sum_{n=0}^{N} {(-1)}^{N}A_n^{(N)}(iB_h(i))^n=\sum_{n=0}^{N} {(-1)}^{N+n}A_n^{(N)}\left(2\pi( h+E_0)-\frac{1}{4}-\frac{2\pi\delta_{h,0}}{e^{2\pi}-1}\right)^{n}.
\end{align}

%
%
 %
 \section{Spin Dependent Bound}\label{secch}
 In this section, we obtain an upper bound on the  scaling dimension $h$ and $\bar{h}$. For deriving an upper bound on $h$ and $\bar{h}$, we consider the medium temperature expansion for $N_R=0$ and $N_L=0$ respectively. 
 First we derive an upper bound on $h$, and the upper bound on $\bar{h}$  is obtained in the same way.

 Now, let us rewrite the partition function \eqref{p2} as follows
 \begin{align}\label{deco}
 Z(\t,\tb)= Z^{0}(\t,\tb)+ Z^A(\t,\tb),
 \end{align} 
 where
  \begin{align}\label{z0}
 Z^{0}(\t,\tb)=\chi _0(\tau)\left(\bar{\chi} _{0}(\bar{\tau})+\sum _{\bar{h}}\rho(\bar{h})\bar{\chi} _{\bar{h}}(\bar{\tau})\right),
 \end{align} 
and
\begin{align}\label{za}
Z^A(\tau,\bar{\tau})&=\sum _{h}\rho(h)\chi _h(\tau)\bar{\chi} _{0}(\bar{\tau})+\sum _{h,\bar{h}}\rho(h,\bar{h})\chi _h(\tau)\bar{\chi} _{\bar{h}}(\bar{\tau})\\\nn
&=\sum _{h,\bar{h}}\rho(h,\bar{h})\chi _h(\tau)\bar{\chi} _{\bar{h}}(\bar{\tau})(1-\delta_{h,0}).
\end{align}
Using medium temperature expansion \eqref{de1} for $N_L\ne 0 ,N_R=0$ and  $N'_L\ne 0 ,N'_R=0$, \eqref{d3} and, decomposition of partition function \eqref{deco}, yields
\begin{align}\label{cons2}
\sum _{h,\bar{h}}\rho(h,\bar{h})g^{N_L}(h+E_0)\chi _h(i)\bar{\chi} _{\bar{h}}(-i)(1-\delta_{h,0})=-g^{N_L}(E_0) Z^{0}(i,-i).
\end{align}
\begin{align}\label{cons3}
\sum _{h,\bar{h}}\rho(h,\bar{h})g^{N'_L}(h+E_0)\chi _h(i)\bar{\chi} _{\bar{h}}(-i)(1-\delta_{h,0})=-g^{N'_L}(E_0) Z^{0}(i,-i).
\end{align}
dividing both side of equations \eqref{cons2} and \eqref{cons3}  leads to
\begin{align}\label{23}
\dfrac{\sum_A \rho(h_{A},\bar{h}_{A})F(h_A)(1-\delta_{h_{A},0})\Lambda_A e^{-2\pi \Delta_A }}{g^{N_L^{\prime}}(E_0)\sum_B g^{N_L^{\prime}}(h_B+E_0)(1-\delta_{h,0}) \Lambda_B e^{-2\pi \Delta_ B }}=0.
\end{align}
where $\Lambda=(1-e^{-2\pi})^{\delta_{\bar{h},0}}$. The summation is over all primary fields and the conformal weight of the primary fields arrange as follows 
\begin{align}\label{24}
0=h_0<h_1\leq h_2\leq h_3 \leq \cdots
\end{align}
and the polynomial $F(h)$ are defined as follows
\begin{align}\label{22}
F(h)=g^{N_L^{\prime}}(E_0)g^{N_L}(h+E_0)-g^{N_L}(E_0)g^{N_L^{\prime}}(E_0+h).
\end{align}

 $N_L$ and $N^{\prime}_L$ are odd, therefore  \eqref{g1-1} shows  $F(h)$ is an  odd polynomial in $h$. Every odd polynomial has at least one real root. Let us denote the largest real root of the polynomial $F(h)$ with $h^+$. 
 
 Without loss of generality, suppose that $N_L > N_L ^{\prime}$. In the limit $h\rightarrow \infty$ the polynomial $F(h)$ goes to infinity as well. Therefore, for $h>h^+$ the polynomial $F(h)$ is positive. So, for $h_1>h_+$, we have the following inequalities
 \begin{align}\label{25}
h_n\geq h_1>h^+,\quad\quad\mbox{ for all}\qu n\geq 1
\end{align}
and
\begin{align}\label{26}
F(h_n)>0.
\end{align}
Therefore, every terms in the numerator is positive.
Now suppose that $\tilde{h}^+$ is the largest real root of the polynomial $g^{N_L^{\prime}}(h+E_0)$. Similarly,  for $h_1>\tilde{h}^+$ every terms in the dominator is positive. Hence, for
\begin{align}\label{h+}
h_1>{\rm max}\,(h^+ , \tilde{h}^+).
\end{align}
All terms in the numerator and denominator of the right hand side of \eqref{23} are positive, which is in contrast with the left hand side of this equation. Therefore, our hypothesis \eqref{h+} is not correct and
\begin{align}\label{h+1}
h_1\leq{\rm max}\,(h^+ , \tilde{h}^+).
\end{align}
In appendix \eqref{apa} the values of the $h^+$ and $\tilde{h}^+$  in large central charge limit  calculated as follows
\begin{eqnarray}\label{h2+}
h^+=\frac{c}{12}-\dfrac{1}{12}-\frac{1}{4\pi}+\dfrac{N_L+N'_L}{2 \pi}.
\end{eqnarray}
\begin{equation}\label{ht+}
\tilde{h}^+=\frac{c}{24}+\mathcal{O}(1),
\end{equation}
using \eqref{h2+} and \eqref{ht+},  the upper bound on the first primary fields $h_1$, in the large central charge limit can be obtained as follows
\begin{align}\label{h1}
h_1\leq\dfrac{c}{12}-  \dfrac{1}{12}-\dfrac{1}{4\pi}+\dfrac{1}{e^{2\pi}-1}+\dfrac{N_L+N'_L}{2 \pi}.
\end{align}
Using the similar method, an upper bound on the $\bar{h}_1$ is obatained as follows
\begin{align}\label{h1}
\bar{h}_1\leq\dfrac{\bar{c}}{12}-  \dfrac{1}{12}-\dfrac{1}{4\pi}+\dfrac{1}{e^{2\pi}-1}+\dfrac{N_R+N'_R}{2 \pi}.
\end{align}
as \eqref{h1} shows,  with increase of the order of derivatives the upper bound does not improve. The best upper bound can be obtained for the minimum value of the $(N_L+N'_L)$. Thus, the best upper bound derived for $N_L+N'_L=4$($N_R+N'_R=4)$:
\begin{align}\label{h11}
h_1=\dfrac{\Delta _1+j_1}{2}<\dfrac{c}{12}+0.47558 .
\end{align}
\begin{align}\label{hb11}
\bar{h}_1=\dfrac{\Delta _1-j_1}{2}<\dfrac{\bar{c}}{12}+0.47558 .
\end{align}
using \eqref{h11} and \eqref{hb11}  upper bound on the scaling dimension of the first primary fields is calculated as follows
\begin{align}\label{delta11}
\Delta _1\leq {\rm min}(\dfrac{c}{6}-j_1+0.95, \dfrac{\bar{c}}{6}+j_1+0.95).
\end{align}
where $-\Delta\leq j\leq\Delta$. As \eqref{delta11} shows, the upper bound depends on the spin of the corresponding primary fields. For theory with  which the first primary field have heavy spin the upper bound is of order $\frac{c}{12}$. If the first primary fields has chiral symmetry $j_1=\Delta_1$, therefore, \eqref{delta11} yields
\begin{align}\label{delta1}
\Delta _1\leq {\rm min}(\dfrac{c}{12}+0.47, \dfrac{\bar{c}}{12}+0.47) .
\end{align}
\subsection{ Upper Bound}
In section \eqref{secch}, for special combinations of derivatives, we have shown that by increasing the order of differential we do not obtain better upper bound. In this section we will show that for every linear combination of derivatives for $N<<\sqrt{c}$, it is not possible to improve the upper bound\footnote{This argument is due to Farhang Loran}.

Let us consider \eqref{cons2} , in the large central charge limit for $N<<\sqrt{c}$ (as we show in appendix\eqref{apa}), we can expand $g^{N_L}(h)=h^{N_L}$, therefore, this equation is reduced to
\begin{align}\label{odd1}
\sum_{A\ne 0}e^{-2\pi\Delta_A}(x_A-1)^{N_L}=u(\bar{h}),
\end{align}
where $x_A=\frac{-h_A}{E_0}$ and
\be
 u(\bar{h})=(1-e^{-2\pi})^2+(1-e^{-2\pi})\sum_A\rho(\bar{h_A})e^{-2\pi\bar{h}_A},
 \ee
 Since \eqref{odd1} holds for every odd value of $N_L$ we conclude
\begin{align}\label{odd2}
\sum_{A\ne 0}e^{-2\pi\Delta_A}f_{\mbox{odd}}(x_A-1)=u(\bar{h}) f_{\mbox{odd}}(1),
\end{align}
where  $f_{\mbox{odd}}(x)=-f_{\mbox{odd}}(-x)$ is a (bounded) odd function on $\mathbb{R}$. Now let us consider
two such functions $f_1$ and $f_2$ and let
\begin{align}
f_3(x) = f_1(x)-\frac{f_1(1)}{f_2(1)}f_2(x).
\end{align}
So $f_3(1) = 0$. For general functions $f_1$ and $f_2$ (i.e., without fine-tuning) $x = 1$
is a simple zero of $f_3(x)$ so it changes sign at $x = 1$. \eqref{odd1} gives
\begin{align}\label{odd3}
\sum_{A\ne 0}e^{-2\pi\Delta_A}f_{3}(x_A-1)=0.
\end{align}
The best upper bound can be obtained if the sign of $f_3(x)$ is constant for $x_A - 1 > 1$, and it is given by $x_1 - 1 < 1$.

\section{ An Upper Bound for Even Spin CFT  }\label{even spin}
In  \cite{Qualls-14}  using the invariance of partition function under $ST$ transformation in the imaginary axis, an upper bound on the lowest primary fields for even spin CFTs obtain as follows
\be\label{evup}
\Delta_1<\frac{c}{12}+0.09280.
\ee
In this section we  release the limitation of partition function in the imaginary axis and study the full upper half plane. As we show the upper bound depends on the spin of the primary field. 
\subsection{Intermediate Temperature Expansion}
Using the decomposition of partition function \eqref{deco}, and the derivative of partition function \eqref{d1}, and \eqref{d22-2}, for CFTs with even spin, the medium temperature expansion   in $\tau=i$ and  the intermediate temperature expansion in $\tau=\frac{-1}{2}+i\frac{\sqrt{3}}{2}$ for first order derivative  leads to 
\begin{align}\label{es7}
\sum _{A}B_{h_A}(i)e^{-2\pi\Delta _A}(1-e^{-2\pi})^{\delta _{\bar{h}_A,0}}(1-\delta_{h,0})=B_{0}(i)(1-e^{-2\pi})K(i)
\end{align}
\begin{align}\label{es6}
\sum _{A}B_{h_A}(\frac{-1}{2}+i\frac{\sqrt{3}}{2})e^{-\pi \sqrt{3}\Delta_A}(1+e^{-\pi\sqrt{3}})^{\delta _{\bar{h}_A,0}}(1-\delta_{h,0})=B_{0}(\frac{-1}{2}+i\frac{\sqrt{3}}{2})(1+e^{-\pi\sqrt{3}})K(\frac{-1}{2}+i\frac{\sqrt{3}}{2}).
\end{align}
Where
\be
K(x)=1-e^{2i\pi x}+\sum_{\bar{h}}\rho(\bar{h})e^{2\pi ix}
\ee
Now by dividing  the two equations \eqref{es6} and \eqref{es7}, we obtain
\begin{align}\label{es8}
\frac{\sum _{A}Y(h_A)(1+e^{-\pi\sqrt{3}})^{\delta _{\bar{h}_{A,0}}}(1-\delta_{h,0})e^{-\pi \sqrt{3}\Delta_A}}{\sum _{B}B_{h_B}(\frac{-1}{2}+i\frac{\sqrt{3}}{2})e^{-\pi \sqrt{3}\Delta_A}(1+e^{-\pi\sqrt{3}})^{\delta _{\bar{h}B,0}}(1-\delta_{h,0})}=0,
\end{align}
where
\begin{align}\label{es9}
Y(h):=B_{h}(i)e^{-\pi\alpha\Delta }(1-e^{2\pi}/1+e^{-\pi\sqrt{3}})^{\delta _{\bar{h},0}}-W(E_0)\frac{K(i)}{K(\frac{-1}{2}+i\frac{\sqrt{3}}{2})}B_{h}(\frac{-1}{2}+i\frac{\sqrt{3}}{2}),
\end{align}
and
\begin{align}\label{es10}
W(E_0):=\frac{B_{0}(i)(1-e^{-2\pi})}{B_{0}(\frac{-1}{2}+i\frac{\sqrt{3}}{2})(1+e^{-\pi\sqrt{3}})},
\end{align}
and
\begin{align}
\alpha =2-\sqrt{3}.
\end{align}
Suppose that $h^+_1$ is the largest real root of $Y(h)$, and $\tilde{h}^+_1$  is the largest real root of $B_h(\frac{-1}{2}+i\frac{\sqrt{3}}{2})$ . In order to obtain an upper bound we will use contradiction. For $h>h^+_1$ the polynomial $Y(h)$ is positive. Hence, for 
\begin{align}\label{es11}
h_n>h_1>h^+_1
\end{align}
 all terms in the numerator of equation \eqref{es8} are positive. Similarly for $h>\tilde{h}^+_1$all terms of dominator are positive. Hence, for $h_1>{\rm max}\, (h_1^+ , \tilde{h}_1^+)$ all terms of numerator and denominator are positive, which is in contrast with equation  \eqref{es8}  . We conclude
\begin{align}\label{es12}
h_1\leq {\rm max}\, (h_1^+ , \tilde{h}_1^+),
\end{align}
Using the \eqref{d22-2} value of $\tilde{h}_1^+$, obtain as follows
\begin{align}\label{es13}
\tilde{h}_1^+=\frac{c}{24}-\frac{1}{24}+\frac{\sqrt{3}}{12\pi}\approx \frac{c}{24}+0.004.
\end{align}
In the next section, we  obtain the value of $h^+_1$.

\subsection{The Value of $h^+_1$ }
In order to obtain the value of $h^+_1$, we first  rewrite the polynomial $Y(h)$ in terms of $H^+=2\pi(h+E_0)$:
\begin{align}\label{es14}
Y(H^+)&=[H^+-\frac{1}{4}]\left(1-e^{-2\pi}/1+e^{-\pi\sqrt{3}}\right)^{\delta _{\bar{h}_A,0}}e^{-\pi\alpha\Delta}\\\nonumber
&-\frac{K(i)}{K(\frac{-1}{2}+i\frac{\sqrt{3}}{2})}W(E_0)[H^+-\frac{\sqrt{3}}{6}]=0,
\end{align}
\begin{figure}[htbp]\label{W-1}
\begin{center}
  \includegraphics[width=0.7\textwidth ,height=0.3\textheight]{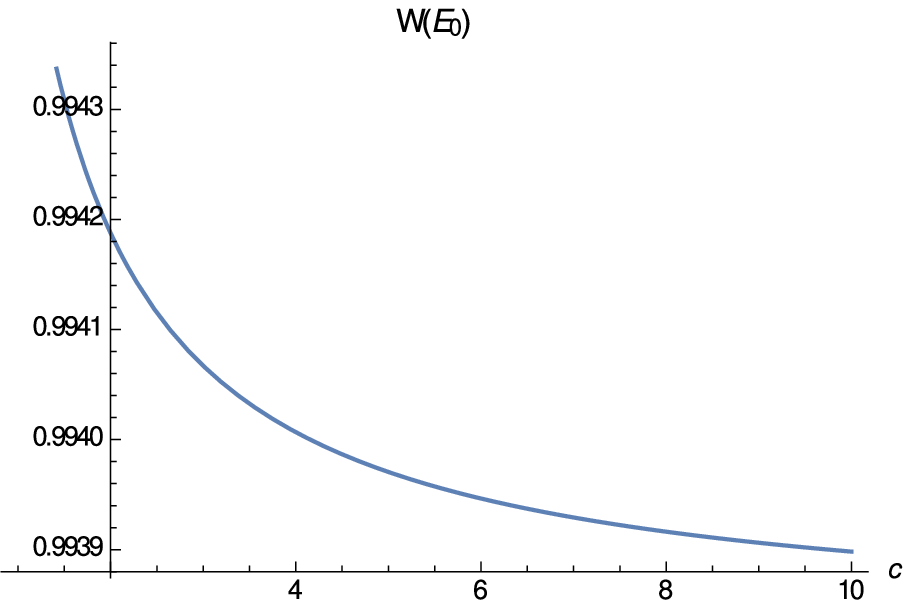}\\
  \caption[] { { The value of $W(E_0)$ vs  central charge }}
   \label{figure:w1}
 \end{center}
 \end{figure}  
\begin{figure}[htbp]\label{Q-1}
 \begin{center}
  \includegraphics[width=0.7\textwidth ,height=0.3\textheight]{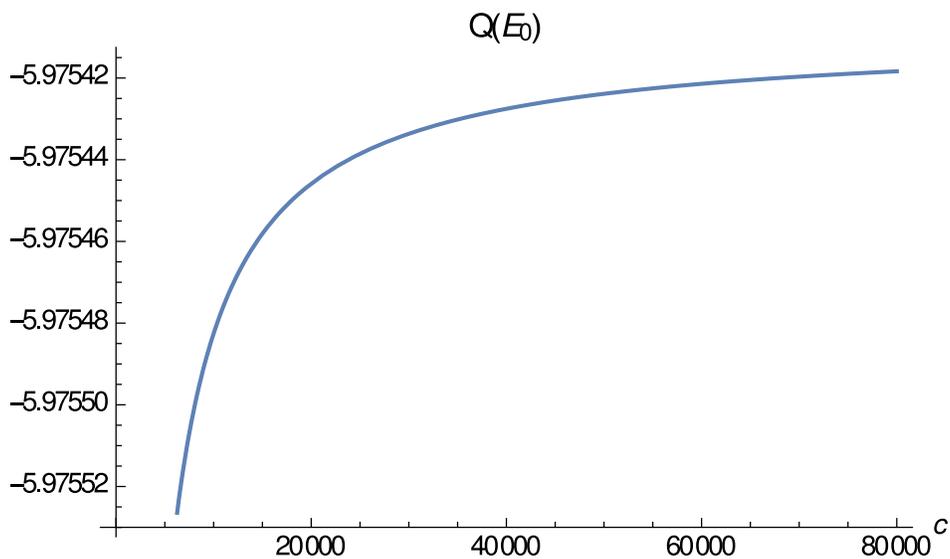}\\
  \caption[] { { The value of $Q(E_0)$ vs  central charge }}
 \label{figure:q1}
 \end{center}
 \end{figure}  
Figure \eqref{figure:w1} shows the function $W(E_0)$ in terms of central charge $c$. As this figure shows for $c>1$ the function $W(E_0)$ is always positive. Therefore, the equation \eqref{es14} has solutions when both terms in the bracket have the same sign  where it happens for $H^+<\frac{1}{4}$ and $H^+>\frac{\sqrt{3}}{6}$.

For $H^+>\frac{\sqrt{3}}{6}$, since $H^+$ is positive from equation \eqref{es14}  we have
 \begin{align}\label{es17}
&W(E_0)[H^+-\frac{\sqrt{3}}{6}]\\\nonumber
&=[H^+-\frac{1}{4}]\frac{K(\frac{-1}{2}+i\frac{\sqrt{3}}{2})}{K(i)}(1-e^{-2\pi}/1+e^{-\pi\sqrt{3}})^{\delta _{\bar{h}_A,0}}e^{ -\pi\alpha\Delta}<[H^+-\frac{1}{4}]\\\nonumber
&\Rightarrow H^+<Q(E_0),
\end{align}
where
\begin{align}\label{es18}
Q(E_0):=\frac{\frac{\sqrt{3}}{6}W(E_0)-\frac{1}{4}}{W(E_0)-1}.
\end{align} 
As figure \eqref{figure:q1} shows, the function $Q(E_0)$ is a decreasing function of central charge $c$.

For $H^+<\frac{1}{4}$,  equation\eqref{es14} trivially holds and for $H^+>\frac{\sqrt{3}}{6}$  this equation also hold. Therefore, in both cases we have
\begin{align}\label{es19}
h_1^+&<\frac{c}{24}-\frac{1}{24}+\frac{\frac{\sqrt{3}}{6}W(E_0)-\frac{1}{4}}{2\pi(W(E_0)-1)}\\\nonumber
&<\frac{c}{24}-\frac{1}{24}+\frac{\sqrt{3}}{12\pi}+\frac{0.006}{W(E_0)-1}.
\end{align}
Similarly, by taking the derivatives with respect to $\tb$, we have proven that
\begin{align}\label{es19}
\bar{h}_1^+&<\frac{\bar{c}}{24}-\frac{1}{24}+\frac{\frac{\sqrt{3}}{6}W(\bar{E}_0)-\frac{1}{4}}{2\pi(W(\bar{E}_0)-1)}\\\nonumber
&<\frac{\bar{c}}{24}-\frac{1}{24}+\frac{\sqrt{3}}{12\pi}+\frac{0.006}{W(E_0)-1}.
\end{align}
From these bounds  spin dependent bound for the lowest primary field is obtained:
\begin{align}\label{es22}
\Delta_1<{\rm min}(\frac{c}{12}-j-\frac{1}{12}+0.09, \frac{\bar{c}}{12}+j-\frac{1}{12}+0.09) .
\end{align}
Where for theories which the first primary fields have chiral symmetry, we have
\begin{align}\label{es23}
\Delta_1<{\rm min}(\frac{c}{24}-\frac{1}{24}+0.04, \frac{\bar{c}}{24}-\frac{1}{24}+0.04) .
\end{align}
\section{Conclusion}\label{summery}
In this paper we have studied the modular invariance of grand canonical partition function for UC-CFT with $c,\bar{c}>1$, without any limitations on the chiral algebra of theory.  Using modular bootstrap for an arbitrary order of derivative, we derived the spin dependent upper bound on the scaling dimension of the lowest primary field of the theory. We have shown that by increasing the order of the derivative for $N<\sqrt{c}$, the upper bound does not improve.

 For theory which the first primary fields  have large spin $\frac{j}{\Delta}\approx 1$, the upper bound has improved to
\be
\Delta<{\rm min}(\frac{c}{12}+0.47,\frac{\bar{c}}{12}+047),
\ee

From holographic point of view, 2d extremal CFT are dual to 3d pure   gravity. We derive a bound on the spectrum of the black hole for 3d gravity. We have shown that if the lowest primary fields have large spin, in  the gravity, it corresponds to the excistence of extremal or near extremal BTZ black hole.

For $c=24k$, the partition function has an expansion  in terms of Klein function $J$\cite{appostal}, where for $k=1$ extremal CFTs are known\cite{flm1, flm2}. While for the other value of central charges the existence is not clear\cite{gaber1}.  The CFTS which first primary fields have chiral algebra are a candidate of extremal CFTs, which can be dual to the pure gravity. Therefor, the first primary field above the vacuum in CFT are dual to BTZ black hole. 

We have studied  even spin CFTs. The upper bound for even spin CFTs obtained as $\frac{c}{12}$ \cite{Qualls-14}. Here we obtained  spin dependent upper bound, and show that if the lowest primary fields have chiral algebra the upper bound is improved  to $\frac{c}{24}$. Therefore, we have shown that   there is a relation between the upper bound and the symmetries which exist in the theory, and theories which have more symmetry have better upper bound.

 \subsection*{Acknowledgments}
     I am grateful to Farhang Loran, and Mukund Rangamani for reading the  manuscript and for their useful comments. I would like to thank James P. Crutchfield for his kind support and hospitality in University of California Davis.
\appendix
\section{The large central charge limit}\label{apa}
Let us consider $h^+$ as the largest real root of 
\begin{equation}\label{g3}
F(h)=0.
\end{equation}
 In the large central charge limit,  one  can expand it as  follows
 \begin{align}\label{delta3}
 h^+= \sum _{a=-1} ^{\infty} \delta_{-a} (\dfrac{c}{24})^{-a}.
 \end{align}
 Putting \eqref{delta3} in \eqref{g3}, leads to the polynomial with an arbitrary order in $\frac{1}{c}$. Now, let us assume that the derivative are of order  $\left(\frac{c}{24}\right)^{\alpha_N}$, where  $\alpha _N\ll\frac{1}{2}$ . We will explain the reason for this selection at the end of this appendix.
 Remaining the terms up to leading order in \eqref{g3} yields                                                   
 \begin{align}
 (\delta_1-1)^{N_L} - (\delta_1-1)^{N_L^{\prime}}=0 .
 \end{align} 
The real solutions of this equation are $\delta_1=1,2$. Since $h^+$  is the largest real root of above equation, therefore, $\delta_1=2$ . By fixing $\delta_1=2$  and keeping terms in \eqref{g3} up to order $(\dfrac{c}{12})^{N_L+N^{\prime}_L-1}$, $\delta_0$ can be  obtained as follows
 \begin{align}\label{delta00}
\delta _0 = \dfrac{-1}{12}-\dfrac{1}{4\pi}+\dfrac{1}{e^{2\pi}-1}+\dfrac{N_L+N'_L}{2 \pi}.
\end{align}

Now, we explain the allowed order of derivative. From \eqref{A11} one can show that
\begin{equation}
A_n^N=N^{2n}+\mathcal{O}(N^{2n-1})
\end{equation}
Using the expansion of  $h^+$ \eqref{delta3},  and \eqref{g1-1} one can verify that to leading order in $ \frac{c}{24} $ we have
  \begin{align}\label{ddelta2-2}
g^{(N)}(h^{+}+E_0)=\sum_{n=0}^{N}\left(\frac{\pi c}{12}\right)^{N-n+2n\alpha _N}+\mathcal{O}\left(\frac{\pi c}{12}\right)^{N-n+2n\alpha _N+\alpha _N-1}.
\end{align}
for  $\alpha_N >\frac{1}{2}$ with increasing $n$ the power of $\frac{c}{12}$ in the above polynomial increase. In order to expand  $g^{(N)}(h^+ +E_0)$  in terms of $\frac{1}{c}$   we assume that
\begin{equation}
\alpha_N <<\frac{1}{2}.
\end{equation}        

\end{document}